\documentclass[12pt]{article}
\usepackage{amssymb}
\usepackage{bbm}
\usepackage{epsfig}
\usepackage{array}
\usepackage{float}
\usepackage{dsfont}
\usepackage{amstext}
\usepackage{psfrag}
\usepackage{a4}
\usepackage{a4wide}
\def\be{\begin{equation}}
\def\ee{\end{equation}}
\def\bea{\begin{eqnarray}}
\def\eea{\end{eqnarray}}

\begin{document}

\begin{center}
\baselineskip 20pt 
{\Large\bf Simplified Smooth Hybrid Inflation in 
	\\ \vspace{0.2cm} Supersymmetric SU(5)}

\vspace{1cm}

{\large 
Mansoor Ur Rehman\footnote{Email: \texttt{mansoor@qau.edu.pk}} and Umer Zubair\footnote{Email: \texttt{umer.hep@gmail.com}} 
} 
\vspace{.5cm}

{\baselineskip 20pt \it
Department of Physics,  \\
Quaid-i-Azam University, Islamabad 45320, Pakistan. \\
\vspace{2mm} }

\vspace{1cm}
\end{center}

\begin{abstract}
A scheme of simplified smooth hybrid inflation is realized in the framework of supersymmetric $SU(5)$. The smooth model of hybrid inflation provides a natural solution to the monopole problem that appears in the breaking of $SU(5)$ gauge symmetry. The supergravity corrections with nonminimal K\"ahler potential are shown to play important role in realizing inflation with a red-tilted scalar spectral index $n_s <1$, within Planck's latest bounds. As compared to shifted model of hybrid inflation, relatively large values of the tensor-to-scalar ratio $r \lesssim 0.01$ are achieved here, with nonminimal couplings $-0.05 \lesssim \kappa_S \lesssim 0.01$ and $-1 \lesssim \kappa_{SS} \lesssim 1$ and the gauge symmetry-breaking scale $M \simeq (2.0  -   16.7)  \times 10^{16}$ GeV.
\end{abstract}

%
\section*{\large{\bf Introduction}}%

Supersymmetric (SUSY) hybrid inflation \cite{Dvali:1994ms}-\cite{Rehman:2009yj} offers a natural framework to realize inflation within grand unified theories (GUTs). The GUT symmetry breaking in these models is associated with inflation in accordance with the cosmological observations on the cosmic microwave background (CMB) radiation. In the standard version of SUSY hybrid inflation \cite{Dvali:1994ms,Copeland:1994vg} GUT symmetry is broken at the end of inflation whereas in the shifted \cite{Jeannerot:2000sv} and the smooth \cite{Lazarides:1995vr} versions of the model it is broken during inflation. The copious formation of topological defects (such as magnetic monopoles) are, therefore, naturally avoided in the shifted and smooth versions of SUSY hybrid inflation.

The supersymmetric $SU(5)$ is the simplest GUT gauge symmetry where all three gauge couplings of Standard Model (SM) gauge group, $G_{\text{SM}} \equiv SU(3)_C \times SU(2)_L \times U(1)_Y$, unify at GUT scale $M_G \simeq 2 \times 10^{16}$ GeV. The $SU(5)$ gauge symmetry is usually broken into SM by the non-vanishing vacuum expectation value (VEV) of  {\bf 24} adjoint Higgs field. This then leads to the overproduction of catastrophic magnetic monopoles in conflict with the cosmological observations. To avoid this problem shifted and smooth variants of SUSY hybrid inflation can be employed{\footnote{See ref.~\cite{Antusch:2010va} for an alternative scheme to avoid monopole problem in SUSY GUTs.}}. In shifted hybrid inflation, the addition of leading-order nonrenormalizable term in the superpotential generates an additional classically flat direction. With the necessary slope provided by the one-loop radiative corrections, this shifted track can be used for inflation; see ref.~\cite{Khalil:2010cp} for details. As $SU(5)$ gauge symmetry is broken in the shifted track, catastrophic magnetic monopoles are inflated away. In the smooth version of hybrid inflation, this additional direction appears with a slope at classical level which then helps to drive inflation. The radiative corrections are usually assumed to be suppressed in smooth hybrid inflation. Furthermore, inflation ends in the smooth hybrid model due to slow-roll breaking whereas in standard and shifted hybrid models, termination of inflation is abrupt followed by a waterfall.

In this paper we consider the simplified version of smooth hybrid inflation in SUSY $SU(5)$. In simplified smooth hybrid inflation \cite{Rehman:2012gd}, the ultraviolet (UV) cutoff $M_*$ is replaced with the reduced Planck mass, $m_P \simeq 2.4 \times 10^{18}$ GeV, in contrast to the standard smooth hybrid inflation \cite{Lazarides:1995vr} where $M_*$ is allowed to vary below $m_P$. With the minimal K\"ahler potential, inflation requires trans-Planckian field values which is inconsistent with the supergravity (SUGRA) expansion. However, with the nonminimal K\"ahler potential, successful inflation is realized with sub-Planckian field values. The predicted values of the scalar spectral index $n_s$ and the tensor-to-scalar ratio $r$ are easily obtained within the Planck's latest bounds \cite{Ade:2015xua}. As compared to the shifted case we obtain relatively large values of tensor-to-scalar ratio $r$ in smooth hybrid inflation.

%
%
\section*{\large{\bf Smooth hybrid $SU(5)$ inflation}}%

In SUSY $SU(5)$, the matter superfields are accommodated to the $1$, $\bar{5}$ and $10$ dimensional representations, while the Higgs superfields belong to the fundamental representations $H\,(\equiv 5_{h})$, $\bar{H}\,(\equiv \bar{5}_{h})$ and adjoint representation $\Phi\,( \equiv 24_{H})$.
The superpotential of the model which is consistent with the $R$, $SU(5)$ and $Z_{3}$ symmetries with the leading-order nonrenormalizable terms is given by {\footnote{For SUSY SU(5) inflation without R-symmetry, see ref.~\cite{Covi:1997my}.}}
\bea%
W &=& S\left( \mu^2 + \frac{Tr(\Phi^3)}{m_{P}}
\right) + \gamma \frac{\bar{H} \Phi^3 H}{m_{P}^{2}} + \delta \bar{H} H  \nonumber \\
&+& y_{ij}^{(u)}\,10_i\,10_j\,H + y_{ij}^{(d,e)}\,10_i\,\bar{5}_j\,\bar{H}
+y_{ij}^{(\nu)}\,1_i\,\bar{5}_j\,H + m_{\nu_{ij}}\,1_i\,1_j ,
\eea %
where $S$ is a gauge singlet superfield, $\mu$ is a superheavy mass and the UV cutoff $m_{P}$ (reduced Planck mass) has replaced the cutoff $M_*$ \cite{Rehman:2012gd} usually employed in smooth hybrid inflation models \cite{Lazarides:1995vr}. In the superpotential above, $y_{ij}^{(u)}$, $ y_{ij}^{(d,e)}$, $y_{ij}^{(\nu)}$ are the Yukawa couplings for quarks and leptons, and $m_{\nu_{ij}}$
is the neutrino mass matrix. The first term in the first line of Eq.~(1) is relevant for inflation, while the last two terms are required for the solution of the doublet-triplet splitting problem, for which a fine-tuning on the parameters is required. The terms in the second line generate fermion masses after the electroweak symmetry breaking. The global $U(1)_{R}$ symmetry ensures the linearity of the superpotential $W$ in $S$-omitting terms such as $S^2$ which could spoil inflation by generating an inflaton mass of Hubble size, $H \simeq \sqrt{\frac{\mu^4}{3\,m_P^2}}$. Also, $W$ respects a $Z_{3}$ symmetry under which $\Phi \rightarrow e^{2 \dot{\iota} \pi / 3} \Phi$ and, hence, only cubic powers of $\Phi$ are allowed. All other fields are neutral under the $Z_3$ symmetry. The $R$-charges of the superfields are assigned as follows \cite{Khalil:2010cp}:
\be %
R: S(1), \Phi(0), H(2/5), \bar{H}(3/5), 10(3/10), \bar{5}(1/10), 1(1/2).
\ee %
In component form, the above superpotential takes the following form,%
\be%
W \supset S\left( \mu^2 + \frac{1}{4 m_P} d_{ijk}\phi_{i}\phi_{j}
\phi_{k}\right)+\delta \bar{H_{a}}H_{a} + \gamma \frac{\bar{H}_{a} H_{d}}{m_{P}^{2}} T^{i}_{ab} T^{j}_{bc} T^{k}_{cd} \, \phi_{i} \phi_{j} \phi_{k},%
\label{superpot-shift}%
\ee %
where we have employed the $SU(5)$ adjoint basis for $\Phi = \phi_i T^i$ with Tr$[T_i T_j] = \frac{1}{2}\delta_{ij}$ and $d_{ijk} = 2$Tr$[T_i\{T_j,T_k\}]$. Here the indices $i$, $j$, $k$ vary from 1 to 24 whereas the indice $a$, $b$, $c$, $d$ vary from 1 to 5. The $F$-term scalar potential obtained from the above superpotential is given by %
\begin{eqnarray}%
V_F &=&  \left| \; \mu^2 + \frac{1}{4 m_P}d_{ijk}\phi_{i}\phi_{j}
\phi_{k} \; \right|^{2}+\sum_{i}\left| \; \frac{3}{4 m_P}d_{ijk} S \phi_{j} \phi_{k} + 3 \, \gamma \frac{\bar{H}_{a} H_{d}}{m_{P}^{2}} T^{i}_{ab} T^{j}_{bc} T^{k}_{cd} \, \phi_{j} \phi_{k}  \; \right|^{2}\nonumber\\&&
 + \sum_{d}\left|\delta \bar{H_{d}} + \gamma \frac{\bar{H_{a}}}{m_{P}^{2}} T^{i}_{ab} T^{j}_{bc} T^{k}_{cd} \, \phi_{i} \phi_{j} \phi_{k} \right|^{2}+\sum_{d}\left|\delta
H_{d} + \gamma \frac{H_{a}}{m_{P}^{2}} T^{i}_{ab} T^{j}_{bc} T^{k}_{cd} \, \phi_{i} \phi_{j} \phi_{k} \right|^{2},%
\label{scalarpot-shift}
\end{eqnarray}%
where the scalar components of the superfields are denoted by the same symbols as the corresponding superfields. The vacuum expectation values (VEVs) of the fields at the global SUSY minimum of the above potential are given by,
\bea
S^0 =  H_{a}^0  =  \bar{H_{a}^0} =0, \, Tr[(\Phi^0)^3] = d_{ijk}\phi_{i}^0\phi_{j}^0\phi_{k}^0 = -M^3/\sqrt{15} ,
\label{gmin}
\eea 
where $M = \left(4 \sqrt{15} \, \mu^2 m_P \right)^{1/3} $, and the superscript `0' denotes the field value at its global minimum. Using $SU(5)$ symmetry transformation the VEV matrix $\Phi^0 = \phi_i^0 T^i$ can be aligned in the $24$-direction. This implies that $\phi_i^0 = 0, \, \forall \, i \neq 24$ and $\phi_{24}^0 = M $, where $d_{24\,24\,24} = -1/\sqrt{15}$ and $\phi_i^{0*} = \phi_i^0$ have been assumed. Thus the $SU(5)$ gauge symmetry is broken down to SM gauge group $G_{\text{SM}}$ by the non-vanishing VEV of $\phi_{24}^0$ which is a singlet under $G_{\text{SM}}$. The $D$-term scalar potential,
\bea %
V_D = \frac{g^2}{2} \sum_{i} \left( f^{ijk} \phi_j \phi_{k}^{\dagger}  + T^i \left( \left| H_a \right|^2  - \left| \bar{H}_a \right|^2  \right) \right)^2,
\eea %
also vanishes for this choice of the VEV (since $f^{i, 24, 24} = 0$) and for $\vert \bar{H}_a \vert = \vert H_a \vert$. 

The scalar potential in Eq.~(\ref{scalarpot-shift}) can be written in terms of the dimensionless variables%
\be %
x = \frac{|S|}{M}~, ~~~~~~~~~~~~ 
y = \frac{\phi_{24}}{M}~, %
\ee %
as, 
\be %
V = \mu^4 \left( \left(1-y^3\right)^2 +  9 x^2 y^4\right)~.%
\ee %
 This potential is displayed in Fig.~(1) which shows a valley of minimum given in the large $x$ limit by
\be %
 y = - \frac{2 \times 2^{1/3} x^2}{\left(1 + \sqrt{1 + 32 \, x^6}\right)^{1/3}} + \frac{\left(1 + \sqrt{1 + 32 \, x^6}\right)^{1/3}}{2^{1/3}} ~ \approx \frac{1}{6 \, x^2} \;.%
\ee %
This valley of local minimum is not flat and possess a slope to drive inflaton towards SUSY vacuum.
Here we assume special initial conditions for inflation to occur in the valley. However, see the relevant references in \cite{Tetradis:1997kp} for detailed discussion of the fine-tuning of initial conditions in various models of SUSY hybrid inflation. During inflation ($x \gg 1$), the global SUSY potential is given by,
\be %
V \simeq \mu^4 \left( 1 - \frac{1}{432 \, x^6 } \right).%
\ee %
\begin{figure}[t]
	\centering \includegraphics[width=10cm]{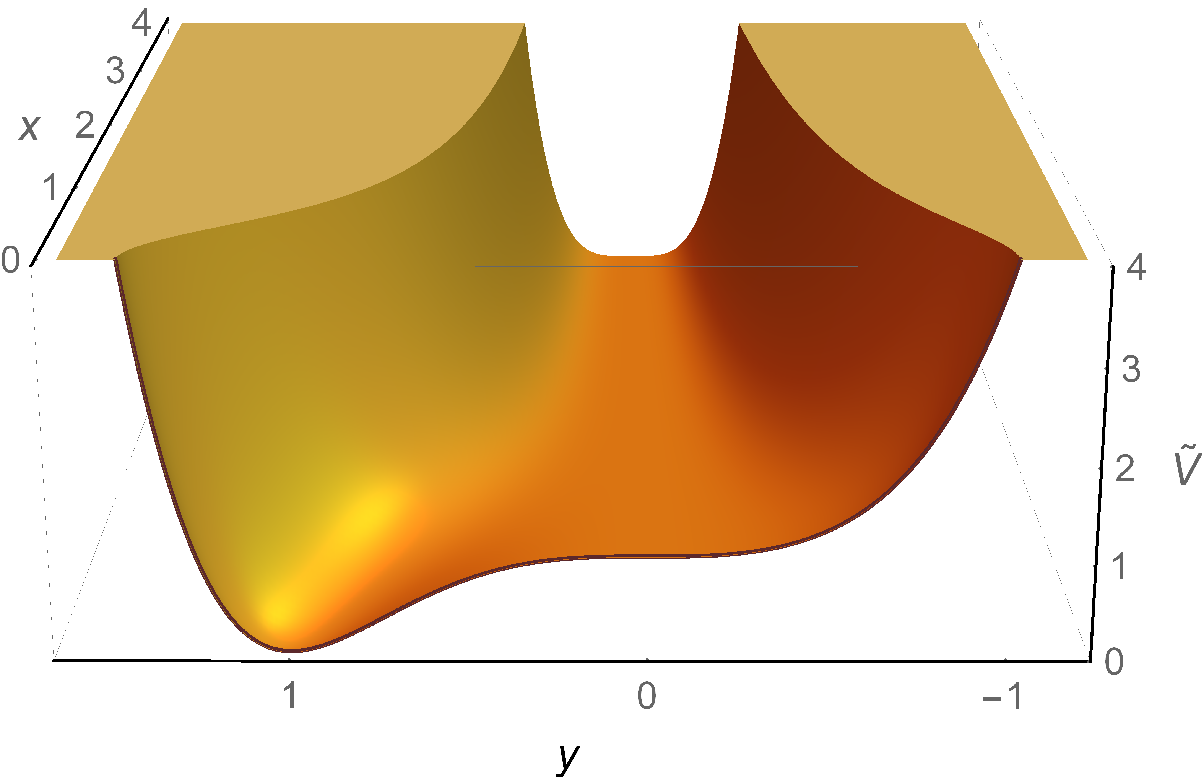}
	\caption{The tree-level, global scalar potential 
		$\tilde{V}=V/\mu^4$ of SUSY $SU(5)$ smooth hybrid inflation.}
	\label{fig1}
\end{figure}
The inflationary slow roll parameters are given by,
\bea
\epsilon = \frac{1}{4}\left( \frac{m_P}{M}\right)^2
\left( \frac{V'}{V}\right)^2, \,\,\,
\eta = \frac{1}{2}\left( \frac{m_P}{M}\right)^2
\left( \frac{V''}{V} \right), \,\,\,
\xi^2 = \frac{1}{4}\left( \frac{m_P}{M}\right)^4
\left( \frac{V' V'''}{V^2}\right). 
\eea
Here, the derivatives are with respect to $x=|S|/M$ whereas the canonically normalized field $\sigma \equiv \sqrt{2}|S|$. In the slow-roll (leading order) approximation, the tensor-to-scalar ratio $r$, the scalar spectral index $n_s$, and the running of the scalar spectral index $dn_s / d \ln k$ are given by
\bea
r &\simeq& 16\,\epsilon,  \\
n_s &\simeq& 1+2\,\eta-6\,\epsilon,  \\
\frac{d n_s}{d\ln k} &\simeq& 16\,\epsilon\,\eta
-24\,\epsilon^2 - 2\,\xi^2 .
\eea
The last $N_0$ number of e-folds before the end of inflation is,
\bea
N_0 = 2\left( \frac{M}{m_P}\right) ^{2}\int_{x_e}^{x_{0}}\left( \frac{V}{%
	V'}\right) dx,
\eea
where $x_0$ is the field value at the pivot scale $k_0$, and
$x_e$ is the field value at the end of inflation, defined by $|\eta(x_e)| = 1$.
The amplitude of the curvature perturbation is given by
\bea \label{perturb}
A_{s}(k_0) = \frac{1}{24\,\pi^2}
\left. \left( \frac{V/m_P^4}{\epsilon}\right)\right|_{x = x_0},
\eea
where $A_{s}= 2.142 \times 10^{-9}$ is the Planck normalization at $k_0 = 0.05\, \rm{Mpc}^{-1}$ \cite{Ade:2015xua}.
In the large $x$ limit we obtain following results for various inflationary parameters, 
\bea
\eta(x_e) &=& -1 \,\,\, \Rightarrow  \,\,\, x_e \simeq \left(\frac{7}{144}\left( \frac{m_P}{M}\right)^2\right)^{1/8}, \\
N_0 &\simeq& 18 \left( \frac{M}{m_P}\right)^2 (x_0^8 - x_e^8) \,\,\, \Rightarrow  \,\,\,
x_0 \simeq \left(\frac{8 N_0}{144}\left( \frac{m_P}{M}\right)^2\right)^{1/8}, \\
A_{s} &\simeq& \frac{18}{5\,\pi^2} \left( \frac{M}{m_P}\right)^8 x_0^{14} 
\,\,\,  \Rightarrow  \,\,\,
\frac{M}{m_P} \simeq \left( \left(18\right)^{3/2} \frac{(5\pi^2)^2 A_s^2}{N_0^{7/2}}\right)^{1/9}, \\
n_s &\simeq& 1 - \frac{7}{72x_0^8} \left( \frac{m_P}{M}\right)^2 \simeq 1 - \frac{7}{4N_0} , \\
r &\simeq& \frac{1}{1296x_0^{14}} \left(\frac{m_P}{M}\right)^2 , \,\,\,
d n_s / d\ln k \simeq -\frac{7}{1296x_0^{16}} \left(\frac{m_P}{M}\right)^4.
\eea
For $N_0 = 50$,
we obtain $n_s \simeq 0.965$, $r \simeq 1.2 \times 10^{-7} $ and  $d n_s / d\ln k \simeq -7 \times 10^{-4}$, with
$x_0 \simeq 3.6$, $x_e \simeq 2.2$ and $M \simeq 2.4 \times 10^{16}$ GeV.
\subsection*{\large{\bf Supergravity corrections and non-minimal K\"ahler
		potential}}

Above analysis is incomplete unless we include SUGRA corrections which have important effect on the global SUSY results. The $F$-term SUGRA scalar potential is given by
\begin{equation}
V_{F}=e^{K/m_P^{2}}\left(
K_{i\bar{j}}^{-1}D_{z_{i}}WD_{z^{*}_j}W^{*}-3 m_P^{-2}\left| W\right| ^{2}\right),
\label{VF}
\end{equation}
with $z_{i}$ being the bosonic components of the superfields $z%
_{i}\in \{S,\Phi,H,\bar{H},\cdots\}$ and where we have defined
\be
D_{z_{i}}W \equiv \frac{\partial W}{\partial z_{i}}+m_P^{-2}\frac{%
	\partial K}{\partial z_{i}}W , \,\,\,
K_{i\bar{j}} \equiv \frac{\partial ^{2}K}{\partial z_{i}\partial z_{j}^{*}},
\ee
and $D_{z_{i}^{*}}W^{*}=\left( D_{z_{i}}W\right)^{*}.$ The K\"{a}hler potential may be expanded as
\bea %
K &=& \vert S \vert^2 + Tr \vert \Phi \vert^2 
+ \vert H \vert^2 + \vert \bar{H}\vert^2  \nonumber \\
&&+ \kappa_{S\Phi} \frac{\vert S\vert^2 \, Tr \vert \Phi \vert^2}{m_P^2}
+ \kappa_{S H} \frac{\vert S \vert^2 \vert H \vert^2}{m_P^2}
+ \kappa_{S \bar{H}} \frac{\vert S \vert^2 \vert \bar{H} \vert^2}{m_P^2} \nonumber \\
&& + \kappa_{H \Phi} \frac{\vert H \vert^2 \, Tr \vert \Phi \vert^2}{m_P^2} 
+ \kappa_{\bar{H} \Phi} \frac{\vert \bar{H} \vert^2 \, Tr \vert \Phi \vert^2}{m_P^2} 
+ \kappa_{H \bar{H}} \frac{\vert H \vert^2 \vert \bar{H} \vert^2}{m_P^2} \nonumber \\
&& + \kappa_S \frac{\vert S\vert^4}{4 m_P^2} 
+ \kappa_{\Phi} \frac{ (Tr \vert \Phi \vert^2)^2}{4 m_P^2} 
+ \kappa_{H} \frac{ \vert H \vert^4}{4 m_P^2} 
+ \kappa_{\bar{H}} \frac{ \vert \bar{H} \vert^4}{4 m_P^2} \nonumber \\
&& + \kappa_{SS} \frac{\vert S\vert^6}{6 m_P^4} 
+ \kappa_{\Phi \Phi} \frac{ (Tr \vert \Phi \vert^2)^3}{6 m_P^4} 
+ \kappa_{H H} \frac{ \vert H \vert^6}{6 m_P^4} 
+ \kappa_{\bar{H} \bar{H}} \frac{ \vert \bar{H} \vert^6}{6 m_P^4}
+ \cdots. %
\label{K}
\eea %
As $\Phi$ is an adjoint superfield, many other terms of the form,
\bea
f\left(\vert S \vert^2, \vert \Phi \vert^2,\,\frac{Tr(\Phi^3)}{m_P} + h.c., \cdots \right) \, ,
\eea
can appear in the K\"ahler potential. The effective contribution of all these terms is either suppressed or can be absorbed into other terms already present in the K\"ahler potential.
Therefore, the supergravity (SUGRA) scalar potential during inflation becomes
\bea %
V_{\text{SUGRA}} = \mu^4 \, \left( 1 - \frac{1}{432 \, x^6} - \kappa_S \, x^2 \left(\frac{M}{m_{P}}\right)^2 + \gamma_S \, \frac{x^4}{2} \left(\frac{M}{m_{P}}\right)^4 + \cdots \; \right),
\eea %
where $\gamma _{S}=1-\frac{7\kappa _{S}}{2}+2\kappa _{S}^{2}-3\kappa _{SS}$. As expected, the dominant contribution in the potential comes only from the terms with higher powers of $S$ as all other fields ($\vert \Phi \vert \sim (M/ \vert S \vert)^2 M$ and $H = 0$) are suppressed as compared to $\vert S \vert \gg M$. The one-loop radiative corrections and the soft SUSY breaking terms are expected to have a negligible effect on the inflationary predictions, therefore, in numerical calculations we can safely ignore these contributions \cite{Rehman:2012gd}.

Including SUGRA corrections usually generates a large inflaton mass of the order of the Hubble parameter which makes the inflationary slow-roll parameter $\eta \sim 1$ and spoils inflation. This is known as the $\eta$ problem \cite{Linde:1997sj}. In SUSY hybrid inflation with minimal K\"ahler potential, this problem is naturally resolved as a result of a cancellation of the mass squared term from the exponential factor and the other part of the potential in Eq. (\ref{VF}). This is a consequence of $R$-symmetry {\footnote{See ref.~\cite{Civiletti:2013cra}, for a SUSY hybrid inflation scenario (with minimal K\"ahler potential) in which the $R$ symmetry is explicitly broken by Planck scale suppressed operators in the superpotential.}} which ensures this cancellation to all orders \cite{Copeland:1994vg, Linde:1997sj}. With non-minimal K\"ahler potential, however, a mass squared term of the form,
\bea
\sim \; \kappa_S \; H^2 \, |S|^2 \;,
\eea
appears which requires some tuning of the parameter $\kappa_S$ ($\kappa_S \lesssim 0.01$), so that the scalar potential is flat enough to realize successful inflation. 

 It can readily be checked that, for the minimal K\"ahler potential (with $\kappa_S = \kappa_{SS} = 0$), the SUGRA corrections dominate the global SUSY potential for the values $x_0 \simeq 3.6$ and $M \simeq 2.4 \times 10^{16}$ GeV obtained earlier. This, in turn, alter the values of $n_s$ and $r$ significantly, making their predictions lie outside the Planck data bounds. Stating the same fact in a different way, the SUGRA corrections require trans-Planckian field values to obtain $n_s$ and $r$ within Planck's bounds. But this invalidates the SUGRA expansion itself. The minimal case ($\kappa_S = \kappa_{SS} = 0$) is, therefore, inconsistent with the Planck's data. 
 
For non-minimal K\"ahler potential, we obtain the following approximate results for $n_s$ and $r$ in the large $x$ limit,
\bea
n_s &\simeq& 1 \;-\; 2 \kappa_S \;+\; \left(-\frac{7}{72 \, x_{0}^{8}} \;+\; 6 \, \gamma_S \, x_{0}^2  \left(\frac{M}{m_P}\right)^4 \right) \left(\frac{m_P}{M}\right)^2, \\
r &\simeq& 4 \; \left(\frac{1}{72 \, x_{0}^{7}} \left(\frac{m_P}{M}\right) \;-\; 2 \, \kappa_S \, x_0 \left(\frac{M}{m_P}\right) \;+\; 2\, \gamma_S \, x_{0}^{3} \left(\frac{M}{m_P}\right)^3 \right)^2.
\eea
Now with the addition of two extra parameters, $\kappa_S$ and $\gamma_S$ we expect to find the red-tilted ($n_s < 1$) solutions consistent with the latest Planck bounds on $n_s$. For example, with $S_0 = m_P$, $n_s=0.968$ and $r=0.01$ we obtain $\kappa_S \simeq -0.045$ and $\gamma_S \simeq -0.02$ from above expressions. This rough estimate guides us to the region of parameters where we can possibly find the large $r$ solutions.

\begin{figure}[t]
	\centering \includegraphics[width=7.6cm]{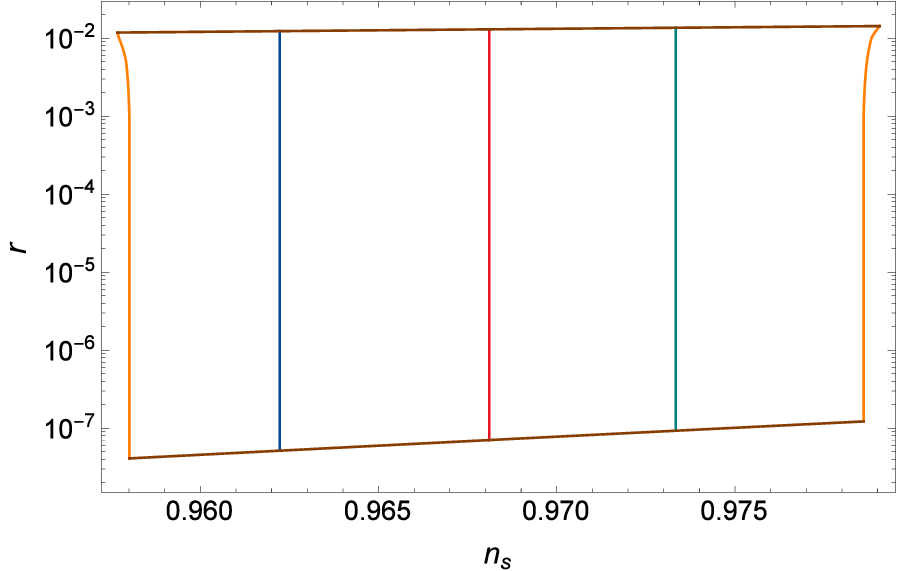}
	\centering \includegraphics[width=7.9cm]{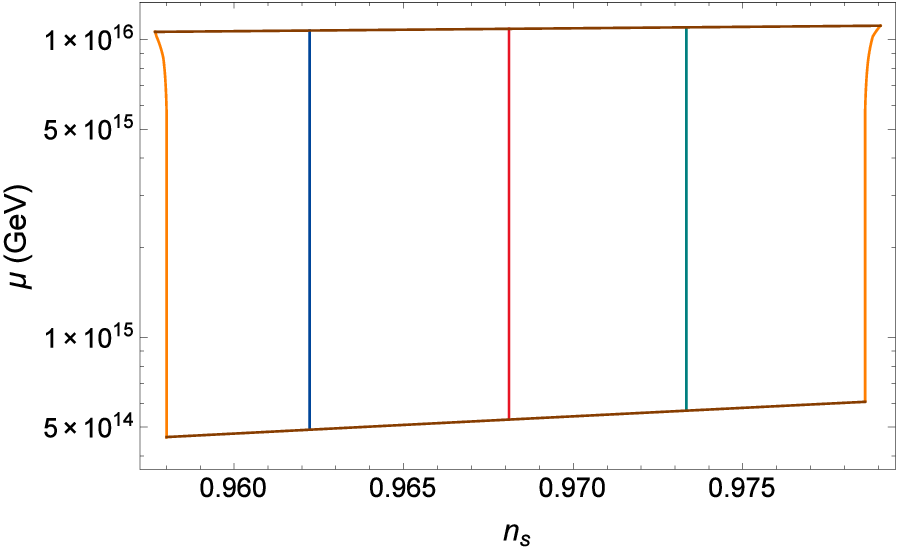}
	\caption{The tensor-to-scalar ratio $r$ and mass scale $\mu$ versus the scalar spectral index $n_s$ with $N_0 = 50$. The orange curves represent the Planck $2$-$\sigma$ bounds, while the upper and lower (brown) curves represent the $\left|S_0\right| = m_P$ and $\kappa_{SS} = -1$ constraints, respectively. The blue, red and green curves are drawn for $n_s = 0.9622$ (1-$\sigma$), $0.9681$ (central value) and $0.9733$ (1-$\sigma$) all with zero tensor modes ($r=0$), respectively.}
	\label{fig2}
\end{figure}

\begin{figure}[t]
	\centering \includegraphics[width=7.8cm]{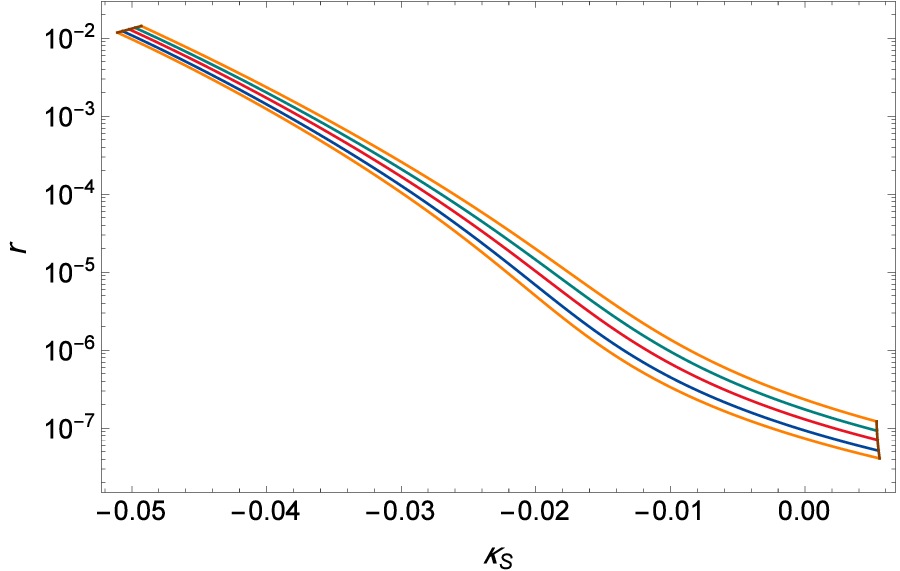} \quad
	\centering \includegraphics[width=7.38cm]{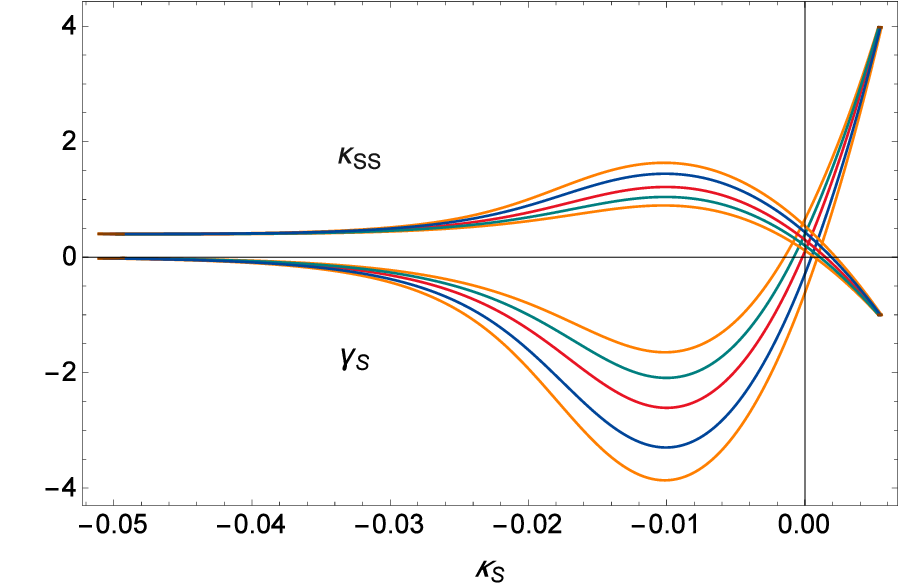}
	\caption{The tensor-to-scalar ratio $r$ and the non-minimal couplings $\gamma_S$ and $\kappa_{SS}$ versus $\kappa_S$ for $N_0 = 50$. The Planck $2$-$\sigma$ bounds are shown by the orange curves, while the brown curves represent the $\left|S_0\right| = m_P$ and $\kappa_{SS} = -1$ constraints. The blue, red and green curves are drawn for $n_s = 0.9622$ (1-$\sigma$), $0.9681$ (central value) and $0.9733$ (1-$\sigma$) all with zero tensor modes ($r=0$), respectively}.
	\label{fig3}
\end{figure}

\begin{figure}[!tbp]
	\centering \includegraphics[width=7.5cm]{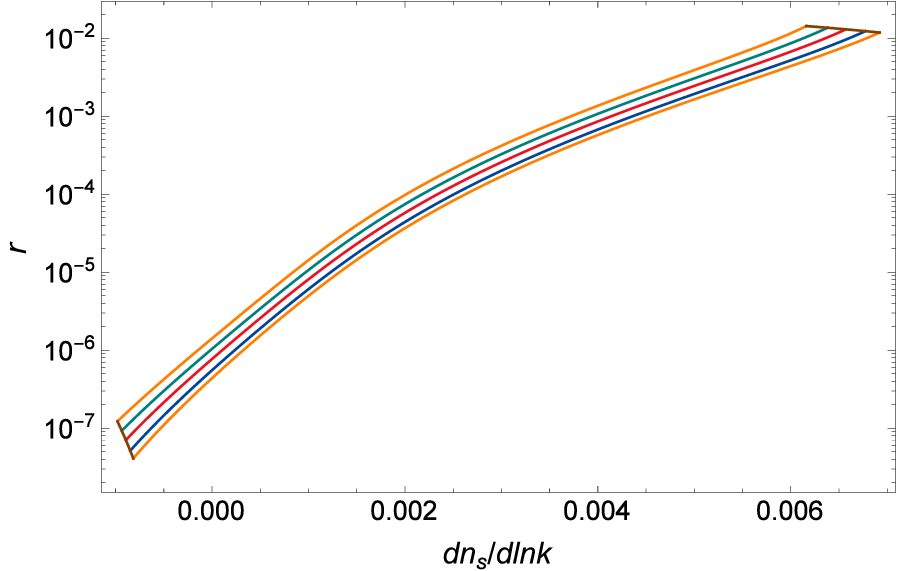}
	\centering \includegraphics[width=7.5cm]{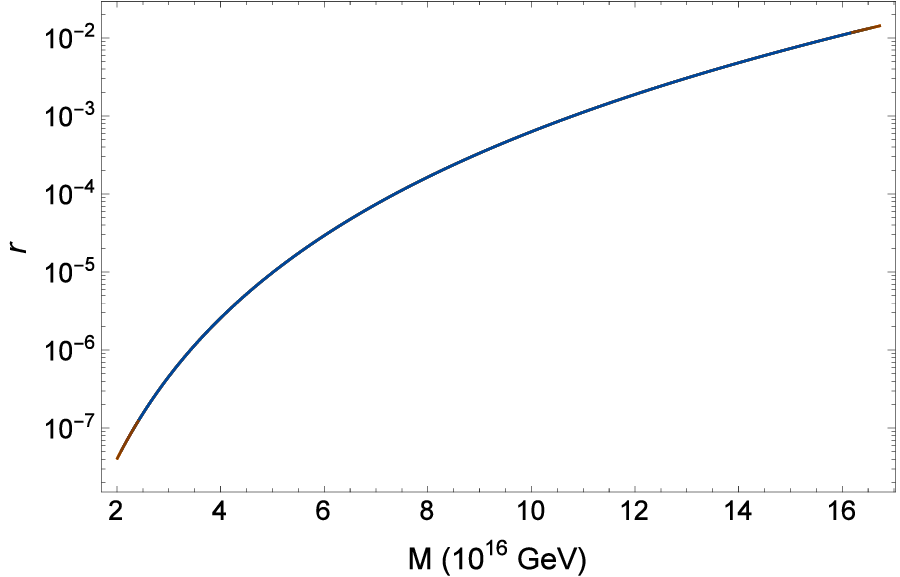}
	\caption{The tensor-to-scalar ratio $r$ versus the running of scalar spectral index $d n_s / d \ln k$ and symmetry breaking scale $M$ for $N_0 = 50$. The orange curves represent the Planck $2$-$\sigma$ bounds, while the upper and lower (brown) curves represent the $\left|s_0\right| = m_P$ and $\kappa_{SS} = -1$ constraints, respectively. The blue, red and green curves are drawn for $n_s = 0.9622$ (1-$\sigma$), $0.9681$ (central value) and $0.9733$ (1-$\sigma$) all with zero tensor modes ($r=0$), respectively.}
	\label{fig4}
\end{figure}

In our numerical calculations we take ($\left|\kappa_S\right|$, $\left|\kappa_{SS}\right|$) $\leq$ 1 and $\left|S_0\right|$ $\leq$ $m_P$. Employing (next-to-leading order) slow-roll approximations \cite{Stewart:1993bc, Kolb:1994ur}, the predicted values of various parameters are displayed in Figs.~(\ref{fig2}-\ref{fig4}). The results obtained here are quite similar to those obtained in the simplified smooth model of hybrid inflation  \cite{Rehman:2012gd}, where instead of adjoint superfield a conjugate pair of chiral superfields ($\Phi, \bar{\Phi}$) is employed. For the predictions of standard, shifted and smooth hybrid inflation with a conjugate pair of chiral Higgs superfields and non-minimal K\"ahler potential see refs.~\cite{BasteroGil:2006cm}-\cite{Civiletti:2011qg}. It can be seen from the Figs.~(\ref{fig2}-\ref{fig4}) that by employing non-minimal K\"ahler potential, there is a significant increase in the tensor-to-scalar ratio $r$. Moreover, both $\kappa_S$ and $\gamma_S$ play the crucial role to bring the scalar spectral index $n_s$ to the central value of Planck data bounds i.e. $n_s \simeq 0.968$, with a large value of tensor-to-scalar ratio $r$ i.e. $r \simeq 0.01$. For $-0.05 \lesssim \kappa_S \lesssim 0.01$ and $-1 \lesssim \kappa_{SS} \lesssim 1$, we obtain scalar spectral index $n_s$ within Planck $2$-$\sigma$ bounds.

The behavior of $r$ and $\mu$ with respect to the scalar spectral index $n_s$ is presented in Fig.~(\ref{fig2}). The relation between these parameters looks very similar which can be understood by noting that $r$ and $\mu$ are proportional to each other. From Eq.~(\ref{perturb}), we obtain the following approximate relation between the tensor-to-scalar ratio $r$ and $\mu$,
\bea
r \simeq \left(\frac{2}{3 \pi^2 A_s}\right) \left(\frac{\mu}{m_P}\right)^4 ,
\eea
which gives an adequate estimate of the numerical results displayed in Fig.~(\ref{fig2}). The upper boundary curve in Fig.~(\ref{fig2}) represents the $|S_0| = m_P$ constraint, whereas the lower boundary curve corresponds the $\kappa_{SS} = -1$ constraint.
The impact of $\kappa_S$ and $\kappa_{SS}$ on the behavior of $r$ is of particular interest. The variation of $\kappa_S$ with respect to $r$ is shown in the left panel of Fig.~(\ref{fig3}), while the right panel shows the relationship between $\gamma_S$, $\kappa_{SS}$ and $\kappa_S$. As depicted in Fig.~(\ref{fig3}), a red tilted ($n_s < 1$) scalar spectral index and a large tensor-to-scalar ratio $r$ require ($\kappa_s$, $\gamma_s$) $<$ ($0$, $0$). Therefore, the large values of $r$ are obtained with the potential of the form,
\bea
\frac{V}{\mu^4} \; = \; 1 + \; \text{Quadratic} \; - \; \text{Quartic}.
\eea
In the right panel of Fig.~(\ref{fig3}), one can see that in the large $r$ limit both $\kappa_S$ and $\kappa_{SS}$ are tuned to make $\gamma_S$ very small.

For the curve depicted in Fig.~\ref{fig4} (right panel), $M$ varies in the range $(2.0 \; - \;  16.7)  \times 10^{16}$ GeV. This shows that small values of $r$ particularly favors $M \sim M_{\text{GUT}}$ ($\simeq 2 \times 10^{16}$ GeV), whereas large tensor-to-scalar ratio $r$ requires $M$ greater than $M_{\text{GUT}}$. The range of running scalar spectral index $d n_s / d \ln k$ is found to be $(-0.001 \;\text{to}\; 0.007)$ (see left panel of Fig.~\ref{fig4}), which is compatible with the Planck's data assumptions on the power-law expansion of $A_s(k)$ \cite{Ade:2015xua}. 

It is worth comparing our results with the model of shifted hybrid inflation \cite{Khalil:2010cp}, where we find the scalar spectral index $n_s \simeq 0.9603$ with a much smaller tensor-to-scalar ratio $r$, taking on values $r \lesssim 10^{-5}$. The reduction in the value of $r$ is mainly due to the dominant contribution of the radiative corrections which keeps the value of $r$ small. This is in contrast to the above model of smooth hybrid inflation where the radiative corrections are suppressed enough to have any influence on the inflationary predictions. However, in shifted model $M = M_{\text{GUT}}$ is easily obtained whereas in above model $M \gtrsim 2.0 \times 10^{16}$ GeV.

After the $SU(5)$ symmetry breaking, the fields $\phi_j$, $j=9,...,20$, acquire super heavy masses, while the fields $\phi_{24}$ and $S$ acquire masses of order $\mu^2 /M$. The octet $\phi_i$, $i = 1,...,8$, and the triplet $\phi_k$, $k = 21$, $22$, $23$, remain massless as shown in \cite{Khalil:2010cp}. The presence of these massless particles (or flat directions) in simple groups like $SU(5)$ \cite{Kyae:2004ft} and $SO(10)$ with a $U(1)_R$ symmetry is a generic feature as pointed out in \cite{Barr:2005xya}. For a relevant discussion also see \cite{Fallbacher:2011xg}. These massless octets and triplets spoil the gauge coupling unification. To preserve unification we can add vector-like fermions as discussed in \cite{Khalil:2010cp}. However, these vector-like fermions do not form a complete multiplet of $SU(5)$  and thus their presence does not respect $SU(5)$ symmetry. In short we have to give up gauge coupling unification in all SUSY $SU(5)$ models of inflation with a $U(1)_R$ symmetry. 

\section*{\large{\bf Summary}}
To summarize, we have analyzed the simplified smooth hybrid 
inflation in supersymmetric $SU(5)$ model. 
As $SU(5)$ gauge symmetry is broken during inflation monopole density is 
diluted and remains under the observable limits.
With minimal K\"{a}hler potential, SUGRA corrections over-dominate all other terms to 
have any inflation. However, with non-minmal terms in the K\"{a}hler potential
successful inflation is realized. We obtain tensor-to-scalar ratio $r\lesssim 0.01$ with
the non-minimal couplings $-0.05 \lesssim \kappa_S \lesssim 0.01$ and 
$-1 \lesssim \kappa_{SS} \lesssim 1$ consistent with the Planck's 2-$\sigma$ bounds on 
the scalar spectral index $n_s$ and tensor-to-scalar ratio $r$. If the detection of gravitational waves is confirmed by Planck's B-mode polarization data then these models will be ruled out.



\end{document}